# Discovering of the L ligand impact on luminescence enhancement of Eu(Dibenzoylmethane)$_3$.L$_x$ complexes employing transient absorption spectroscopy

Stanislav S. Stanimirov[a], Anton A. Trifonov[b], Ivan C. Buchvarov[b]

**Abstract:**

The effect of luminescent enhancement under exchange of the auxiliary ligand in Europium(III) tris(1,3-diphenyl-1,3-propanedionato) monohydrate was investigated by steady-state and time-resolved transient absorption spectroscopy. The excited state relaxation dynamics of this complex was analysedthrough a comparison of the experimental data obtained for several model compounds, namely Eu(DBM)$_3$.NH$_3$, Eu(DBM)$_3$.EDA, Eu(DBM)$_3$.Phen, Al(DBM)$_3$ and dibenzoylmethane (DBM) in various solutions and polymer matrices. The results show there is no linear relationship between enhancement of the emission quantum yield and the luminescent lifetime,which suggests that the auxiliary ligand reducesthe rate of nonradiative relaxation of the lanthanide ion, but also affects the excited state energy transfer from ligand to metal ion.Transient absorption data shows a clear correlation between the efficiency of the energy transfer and the degree of triplet state population expressed by anamplification of the signal for its excited state absorption band on going from Eu(DBM)$_3$.H$_2$O to the Eu(DBM)=.L complex. The results show thatthis auxiliary ligand exchange actsas a "switch" turning theintersystem crossingon or off as a competitive pathway for excited state relaxation of the europium(III) complexes.

*KEYWORDS: Transient absorption, Europium(III) β-diketonates, DBM, Chemosensors.*

[a] Photochemistry group, Faculty of chemistry and pharmacy, Sofia University, 1 J. Bourchier Blvd 1164 Sofia, Bulgaria

[b] Department of Physics Sofia University, 5 J. Bourchier Blvd. 1164 Sofia, Bulgaria

# 1. Introduction

Many products such as sunscreen[1] and organic light emitting devices[2-5] contain dibenzoylmethane (DBM) as a functional compound. The properties of all those products depend on the interaction between DBM moiety and light and its excited state dynamics.

Derivatives of DBM are widely used as UVA absorbers[6, 7] in photoprotective cosmetics due to their characteristic absorption showing long wavelength band to a maximum of around 350 nm.

**CHART 1. Schematic presentation of the energy levels and of the energy transfer from excited ligand to lanthanide ion in Eu(III) complexes.**

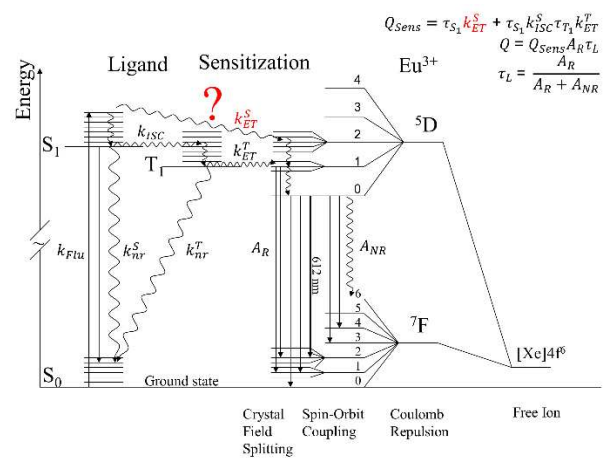

A common use of the β – diketonates is as ligands in many metal complexes[8]. Complexes of lanthanides[5, 9, 10] and iridium ions[11, 12] are employed as emitters or dopands in OLEDs. Some of the most popular compounds with such an application are $Eu(DBM)_3 \cdot L_x$ ternary complexes where the forth (L) ligand is 1, 10-Phenantroline[13], 2, 2'-bipyridine etc[10, 14]. These complexes are photoluminescent and emit red light upon UV irradiation. The characteristic emission is a result of transitions from $^5D_1$ or $^5D_0$ excited states to $^7F_n$ ground state manifold of the $Eu^{3+}$ ion. These transitions are forbidden and direct absorption of the $Eu^{3+}$ ion is unlikely. Photoluminescence occurs through an excitation energy transfer from excited ligand to excited state levels of the metal ion. This is the so-called "antenna effect" whereby DBM or ligand absorbs the light and transfers the excited state energy (electron exchange, the Dexter mechanism) to $Eu^{3+}$ ion. The commonly accepted mechanism for this energy transfer was proposed by Crosby et al. in the early 1960s[15, 16] (CHART 1). Crosby et al. supposed that the excitation energy goes from $S_1$ excited state of the β-diketonate ligand via $T_1$ to $^5D_1$ or $^5D_0$ excited states of the $Eu^{3+}$ ion. Thus, the probability for ISC from singlet to triplet state of the ligand and energy matching between $T_1$ level of the ligand and $^5D_1$ or $^5D_0$ levels of the europium is essential for energy transfer efficiency. This mechanism is still the most commonly proposed in the literature to date[17] and some have applied this to $Eu(DBM)_3 \cdot L_x$ complexes too[5, 18]. According to it, the excited state dynamics of the DBM ligand is crucial for the efficiency of the energy transfer process described above.

**CHART 2. DBM isomers.**

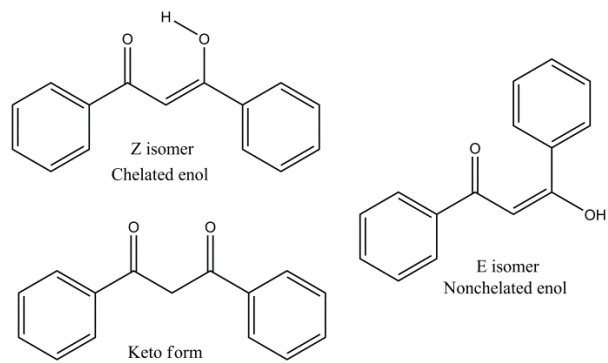

Recent transient absorption studies by Verma et al.[19, 20] report very fast relaxation dynamics of DBM in acetonitrile solution with excited state decay in less than 12 ps. The authors suggest a proton transfer as explanation for the fast dynamics[20]. It is well known that β – diketonates in solution exist in several forms due to keto-enol tautomerization[21-24]. Those are chelated enol (CE), nonchelated enol (NCE) and the keto form (KF) (CHART 2). In the case of DBM the keto-enol equilibrium is shifted nearly 100% to enol form[9]. The results show that the intersystem crossing (ISC) efficiency is very low for DBM in CE form. This constitutes a discrepancy between the proposed energy transfer mechanism and reported excited state dynamic[23, 24] of the DBM CE form at room temperature. Unfortunately, at this point there is no data in the literature about the ISC efficiency of the DBM in its europium complexes, which also possess chelate structure. Most $Eu(DBM)_3 \cdot L_x$ complexes are luminescent, which means that the excited DBM ligand transfers the energy to the $Eu^{3+}$ ion through its triplet state as hypothesised. This raises the question about ISC efficiency in these complexes. Is it reasonable to suggest that the apparent luminescence is evidence for the increased ISC efficiency in these complexes due to the lack of internal conversion to the ground state, as the relaxation path in this bidendate ligand or energy transfer goes through another state and the proposed mechanism is wrong[25]?

These questions were raised after a recent publication, that suggests a possible application of the $Eu(DBM)_3 \cdot H_2O$ complex as a chemosensor for the detection of ammonia[26] and that provoked us to take a look at the DBM complexes again. Mirochnik[26] et al. suggest an explanation of the effect of luminescence enhancement of the polymer films containing $Eu(DBM)_3 \cdot H_2O$ upon ammonia treatment, reported by us[27] almost ten years ago. The authors offerthereduction of the nonradiative pathways as explanation for luminescent intensity enhancement due to the substitution of the water molecules from the inner coordination sphere of the complex with nitrogen-containing ligand.However, our present studiesof the dependence of the $Eu(DBM)_3 \cdot H_2O$ luminescent intensity in ethanol solutionon this substitution suggest something else(see below).

With this study, using transient absorption spectroscopy and steady state absorption and fluorescence spectroscopy, we want to understand the reason for the enhancement of the luminescence of ternary europium complexes upon water substitution and the effect of L ligand on the dynamics of relaxationof several model compounds: Tris (1, 3 - diphenylpropane - 1, 3 - dionato)-europium (III) monohydrate $[Eu(DBM)_3 \cdot H_2O]$; Tris (1, 3 – diphenylpropane - 1,3 - dionato) mono(1, 10 - phenanthroline) europium (III) $[Eu(DBM)_3(Phen)]$; Tris (1, 3 – diphenylpropane - 1, 3 - dionato) (Ethylenediamino) – europium (III) $[Eu(DBM)_3(EDA)]$; Tris (1, 3 – diphenylpropane - 1, 3 - dionato) (ammonio) – europium (III) $[Eu(DBM)_3(NH_3)_x]$ including Dibenzoylmethane [DBM] and Tris(1,3-diphenylpropane-1,3-dionato)-aluminum $[Al(DBM)_3]$, for comparison in various solutions and solid state.

## 2. Experimental

DBM (CAS No. 120–46–7), $Al_2(SO_4)_3 \cdot xH_2O$ (CAS No. 17927-65-0), EDA (CAS No. 107-15-3), $Eu(DBM)_3(Phen)$, Europium(III) tris(1,3-diphenyl-1,3-propanedionato) mono(1,10-phenanthroline) (CAS No. 17904-83-5) was obtained from Sigma-Aldrich and used as obtained. Synthesis of $Eu(DBM)_3(H_2O)$, Europium(III) tris(1,3-diphenyl-1,3-propanedionato) monohydrate is described elsewhere[27]. The $Eu(DBM)_3(EDA)$ complex was obtained by adding an equimolar quantity of EDA toa hot saturated ethanolic solution of $Eu(DBM)_3(H_2O)$ under continuous stirring. After cooling of the solution the yellow precipitate was collected by filtration and washed with absolute ethanol. Elemental analysis: C – 64.07%±0.1%; H – 4.69%±0.1%; N – 3.09%±3%; O – 11.09%±2%. The *in suit*synthesis of the $Eu(DBM)_3(NH_3)$ complexwas made by bubbling ammonia gas through solution of $Eu(DBM)_3(H_2O)$ complex directly in spectroscopic cuvette. The bubbling was stopped when no farther increase in the intensity of the luminescent emission at 612 nm was detected. This complex was not isolated as a separated compound. The ammonia gas was generated by dropping 25% water solution of ammonium hydroxide over pellets of sodium hydroxide. The ammonia gas was passed through a tube filled with pellets of sodium hydroxide. The $Al(DBM)_3$ complex was also synthesized in our laboratory. During continued stirring we added to the water solution of the aluminum sulfate (0.001 mol) an alkaline water solution containing ammonium hydroxide and 0.006 mol DBM. Until the end of the addition the pH of the reaction mixture was kept alkaline. Cooling the flask in an ice bath produced cream colored precipitate forms. The crude crystals were collected by filtration, washed with distilled water and recrystallized from acetone-ethanol mixture. The spectral characterization data for all compounds is given in the supplementary section. Elemental analysis: C – 79.95%; H – 4.76%; O – 14.25%. All the spectroscopic-grade solvents were obtained from Sigma-Aldrich. The NMR spectra were recorded on Bruker Avance III 500 NMR spectrometer in chloroform and acetonitrile solutions. FTIR spectra of the complexes were recorded at Shimadzu FTIR spectrometer 8400S in KBr pellets. The steady-state absorption spectra were recorded on Agilent Cary 5000 UV–VIS-NIR spectrophotometer. The florescence emission and excitation spectra, and the luminescent lifetimes were recorded on Agilent Cary Eclipse fluorescence spectrophotometer. Luminescent quantum yields were determined with Horiba Flurolog 3 series TCSPC spectrophotometer equipped with a quantum yield and a CIE measurement accessory Quanta-phi integration sphere. The polymer films were coated on a glass substrate using Laurel technology WS-400B-6NPP Spin Coater.

Transient absorption measurement were performed on a home-built femtosecond broadband pump-probe setup, virtually identical to one, described previously[28]. The pump wavelength was set to 350 nm for all samples. The changes in optical density were probed by a femtosecond white light continuum (WLC) generated by tight focusing of a small fraction of the output (790nm) of a commercial Ti:sapphire-based pump laser (Integra-C, Quantronix) into a 3-mm calcium fluoride spinning disc. The WLC provides a usable probe source between 320 and 750 nm. The WLC was split into two beams (probe and reference) and focused into the sample using reflective optics. After passing through the sample, both probe and reference beams were spectrally dispersed and simultaneously detected on a CCD sensor. The pump pulse (1 kHz, 300 nJ) was generated by frequency-doubling of the compressed output of a home-built noncollinear optical parametric amplifier system (700 nm, 9 μJ, 40 fs). To compensate for group velocity dispersion in the UV pulse, an additional prism compressor was used. The overall time resolution of the setup was determined by the cross correlation function between pump and probe pulses, which is between 110 and 130 fs (fwhm, assuming a Gaussian line shape). A spectral resolution of ca. 5 nm was obtained for the entire probing range. All measurements were performed with magic angle (54.7°) setting for the polarization of the pump with respect to the polarization of the probe pulse. A sample cell with 1.25 mm fused silica windows and an optical path of 1 mm was used for all measurements. All sample concentrations were adjusted to have similar absorbance of 1 at 350 nm, and pump-probe experiments were conducted at room temperature (ca. 23°C) under continuous shifting of the sample in a plane perpendicular to the laser beams.

Themovement was performed between acquisitions each 80 ms.

The analysis of transient absorption data and graphical presentation of the results was performed by software written by us, which follows the procedure for target analysis and use the theory described here[29, 30].

## 3. Results and discussion

Noticeable enhancement of the red luminescence of the Eu(DBM)$_3$.H$_2$O complex can be detected when trace amounts of ammonium or amines interact with the complex in a solution or in solid state. In this study we try to uncover the reasons for this phenomenon using steady state and time resolved UV spectroscopy and femtosecond transient absorption spectroscopy.We compare the photophysical properties of the Eu(DBM)$_3$.H$_2$O complex with several model compounds which arevarious europium and aluminium complexes of DBM ligand (CHART 3). We use europium derivatives ofthe Eu(DBM)$_3$.H$_2$O complex to finda correlation between luminescenceenhancement and the dynamics of excited state relaxation of the complexes. The aluminiumcomplexes are used to study the dynamicsofexcited state relaxation of the chelated DBM in a compound where UV excitation of the ligand does not lead to any light emission that comes from the energy levels of the metal ion and possessesa mechanism of relaxation different from this presented in (CHART1).

**CHART 3. Structure of the DBM complexes.**

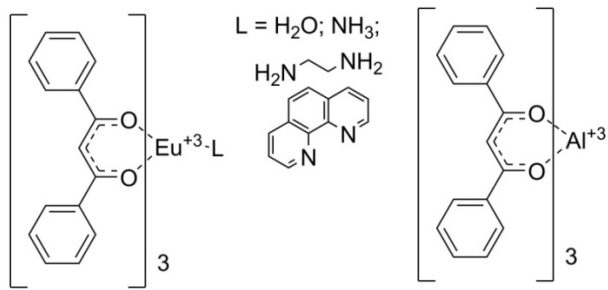

DBM is a light absorber, an antenna which can transfer the excitation energy from its excited states to those of europium (III) ion (CHART1). Following this mechanism, a bright red emission appears from the $^5D_0$ lanthanide ion excited state duringthe relaxation of the Eu(III) ion to the $^7F_j$ ground state manifold.Quantitative measure for the efficiency of this process is overall quantum yield ($Q$), defined astheprobability that the europiumion will emit a photon, given that the ligand has absorbedone.This quantity can be presented as product of couple terms.

$$Q = Q_{sens}.A_R.\tau_L \qquad (1)$$

$Q_{sens}$ is probability excitation energy captured by the ligand to reach $^5D_0$excited state level of the Eu(III) ion.The term $A_R.\tau_L$ is the luminescent quantum yield ($Q_L = A_R.\tau_L$) that gives the probability Eu(III) complex to relax radiativelyfrom its $^5D_0$excited state to the $^7F_j$ ground state manifold, where $\tau_L$isthe lifetime of the $^5D_0$ excited state and $A_R$istherate for spontaneous emissionfrom that state in s$^{-1}$.

Rather than the three bidendate DBM ligands the ternary europium complexes contain another moleculeas auxiliary ligand serving to fill the Eu(III) coordination number. This auxiliary ligand can be a small moleculesuch as water, ammonia or a bigger molecule such as ethylendiamine (EDA) and 1,10 – phenantroline (Phen). Replacing the water molecule from Eu(DBM)$_3$.H$_2$O complex with any of the other nitrogen-containing molecules listed above leads to an apparent increase in the luminescent efficiency of the complex. An explanation of this phenomenon concerning the elimination of the high energy vibrational modes of the water molecules as an effective pathway for non-radiative deactivation of the Eu(III) excited state has been already suggested[26]. However, comparing the change in $^5D_0$ excited state lifetime ($\tau_L$) with the increase of the overall quantum yield ($Q$) of emission no match can be found in the extent of these changes (Table 1). If we assume that only $^5D_0$ excited statelifetime is a function of the type of the auxiliary ligand, then a linear relationship between the overall quantum yield and $\tau_L$is expected,according to Eq. (1). As can be seen in Table 1, a threefoldincrease of the $\tau_L$ going from Eu(DBM)$_3$.H$_2$O ($\tau_L$=0.09 ms) to Eu(DBM)$_3$.EDA ($\tau_L$= 0.26 ms) complex leads to an increase of around 68 times of $Q$ in solid state from 0.73% to 49.5%, respectively. Although a similar observation was made for the samples in solution (Table 1), the change in the values here is much more modest. This means the increase of the overall quantum yieldcannot be explained only by reduction of the $^5D_0$non-radiative relaxation rates.Either ligand-to-metal energy transfer efficiency ($Q_{sens}$) or rates for spontaneous emission from $^5D_0$ state ($A_R$) are also affected.

$$A_R = A\,_{^5D_0 \to\, ^7F_1} n^3 \left(\frac{I_{tot}}{I\,_{^5D_0 \to\, ^7F_1}}\right) \qquad (2)$$

The simplest way to calculate the rates for spontaneous emission of europium complexes is Eq.(2). In this equation the quantitative measure for the $^5D_0$ radiative rates ($A_R$) is the ratio between the integral intensities of the magnetic dipole $^5D_0 \to\, ^7F_1$ transition ($I\,_{^5D_0 \to\, ^7F_1}$) which is independent of the ion's surroundings and thetotal integral intensities ($I_{tot}$) of all transitions with $^5D_0$ origin[31] in the corrected fluorescence emission spectrum of the complexes. This ratio can be used as a qualitative measure of the local symmetry around the europium ion in the complex. That is because$I_{tot}$depends mainly on the band area for the

$^5D_0 \rightarrow {^7F_2}$ transition, which is extremely sensitive to the symmetry of the complex[32, 33]. In this equation $n$ is the refractive index of the medium and $A_{^5D_0 \rightarrow {^7F_1}}$ is spontaneous emission probability for $^5D_0 \rightarrow {^7F_1}$ transition *in vacuo*(calculated[31] to be 14.65 s$^{-1}$).Given that these ratios are virtually equal we calculate almost identical $A_R$ values for Eu(III) complexes in solution (Table 1). This means that replacing a water molecule does not cause a significant change in spontaneous emission rates, so we can expect a significant change in energy transfer efficiency, or $Q_{sens}$. If the refractive index of the medium is known and the quantum yield of the complex in this medium is measured,we can calculate the energy transfer efficiency values (Eq.1 and Eq. 2). The data for Eu(III) complexes in solution are listed inTable 1.

Table 1. Photophysical data for (1) Eu(DBM)$_3$.Phen, (2) Eu(DBM)$_3$.EDA and (3) Eu(DBM)$_3$.H$_2$O complexes.

|  | Solid State | | | Solution (Acetonitrile) | | | Solution (Ethanol) | | |
|---|---|---|---|---|---|---|---|---|---|
| **Complex:** | 1 | 2 | 3 | 1 | 2 | 3 | 1 | 2 | 3 |
| **Temp. (C)** | 23 | 23 | 23 | 20 | 20 | 20 | 20 | 20 | 20 |
| $Q$ | 35.4% | 49.5% | 0.7% | 4.9% | 5.2% | 0.6% | 0.31% | 0.47% | 0.13% |
| $Q_L$ |  |  |  | 16.9% | 14.7% | 4.3% | 3.51% | 3.48% | 4.67% |
| $Q_{sens}$ |  |  |  | 29.1% | 35.5% | 13.9% | 8.83% | 13.51% | 2.78% |
| $\tau_L$ (ms) | 0.40 | 0.26 | 0.09 | 0.25 | 0.22 | 0.07 | 0.07 | 0.07 | 0.07 |
| $A_R$ (s$^{-1}$) |  |  |  | 673 | 671 | 664 | 516 | 527 | 687 |

Q–Overall quantum yield of emission as a product of energy transfer efficiency and luminescent quantum yield, presenting the probability the photons absorbed from ligands to be emitted from excited state of Eu(III) ion; $Q_L$ - Luminescent quantum yield equal to ($\tau_L \cdot A_R$); $Q_{sens}$ - Energy transfer efficiency; $\tau_L$ - Luminescence lifetime; $A_R$ - Radiative rate of the $^5D_0$ state. Standard deviation for all values is below 10% of the value.

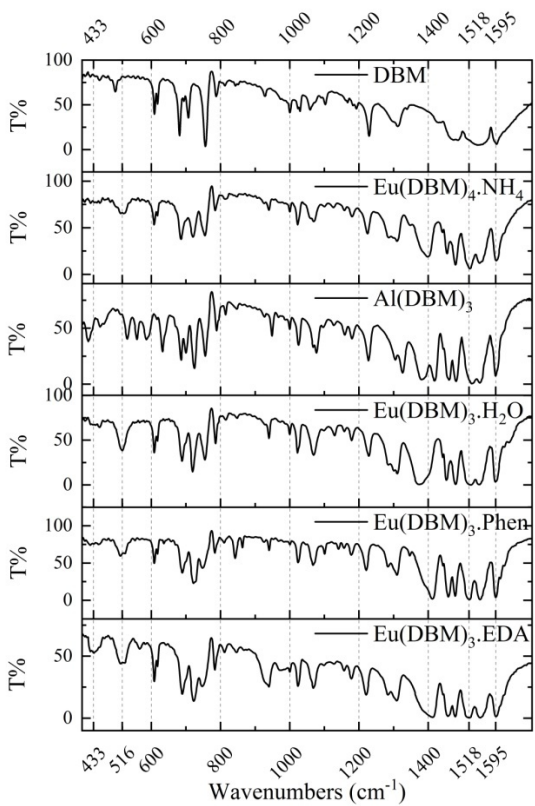

Figure 1 The solid state FTIR spectra of the compounds under study in KBr pellets. The quaternary Eu(DBM)$_4$.NH$_4$ complex is not subject of this publication. The spectrum is shown as supplementary information only for sake of comparison.

### 3.1. Structure of the complexes

From results present in Table 1 it is clear that energy transfer efficiency $Q_{sens}$ increases going from Eu(DBM)$_3$.H$_2$O to Eu(DBM)$_3$.L complexes where L is a nitrogen-containing ligand. $Q_{sens}$ depends on the extent of any relaxation pathway different than this leading to Eu(III) ion excited states and at a distance between the ligand sensitizer and Eu(III) acceptor. Given that energy transfer follows a Dexter mechanism there is an exponential dependence between the efficiency and the distance of ligand to the Eu(III) ion. As a qualitative measure for that distance Eu—O, C=O and C=C vibrations from FTIR spectrum can be used. According to assignments of the IR absorption bands of β-diketonate complexes made by Pinchas[34] et. al. we assign bands at 433 cm$^{-1}$ and 516 cm$^{-1}$ for Eu—O vibrations and bands centered at 1595 cm$^{-1}$ and 1518 cm$^{-1}$ for C=O and C=C vibrations respectively. The shift for all of these bands is insignificant going from a DBM ligand to all of its complexes (Figure 1). Despite the fact that these considerations are indirect, they support the hypothesis that we cannot expect significant changes in chelate. This is even more unlikely in solid state where the effect of luminescent enhancement is dramatic when ammonia vapors are passed over Eu(DBM)$_3$.H$_2$O crystals (Fig. S13). This conclusion is also confirmed by the X-ray data for Eu(DBM)$_3$.H$_2$O and Eu(DBM)$_3$.Phen complexes found in literature[35-38]. There is no significant difference in the length of Eu – O bond which is in the range 2.3±0.1Å. However, the bandcentred at 1400 cm$^{-1}$ inthe IR spectra of the compounds attracts attention (Figure 1). Comparing this band among the FTIR spectra of the compounds shows significantchange. It is clear that this band is doublet and it is best resolved in the case of the Al(DBM)$_3$ complex. Going from the DBM ligand to complexes this band changes its position as well as its intensity. When we compare the IR spectra of the Eu(III) complexes a clear difference inthe position of this band can be detected. The shift of the band going from Eu(DBM)$_3$.H$_2$O to Eu(DBM)$_3$.L complexes is due to the difference in relative intensities of the bands from the doublet. Pinchas[34] et. al. assign this band of acetylacetonate complexes as the asymmetrical C-H bending of the methyl group. Since there are no methyl groups in the DBM compound a revision of the IR bands assignment for these complexes might be needed. We suppose this band also originates from some chelate vibration which is strongly affected by the change in electron density therein. Hypothetically itmight be assigned to C—O stretching of a hydroxyl or an enolate group. Still, a suggestion fora change in electron density in chelate going from Eu(DBM)$_3$.H$_2$O to Eu(DBM)$_3$.L complexes comes after analysis of the NMR spectra where the obvious difference in chemical shifts for αH of DBM ligand can be detected (See NMR spectra in supporting information).

From these results and considerations we can conclude that there is no significant difference in the distance between the sensitizer and the acceptor in any of these compounds, but the increase in electron density around the chelate ring is observed indirectly, by the FTIR and NMR spectra of the complexes, after the replacement of the water molecule from the Eu(DBM)$_3$.H$_2$O with a nitrogen-containing ligand.

These results direct us to look for the answer of this effect on luminescent efficiency in the dynamics of the excited state relaxation of the complexes.

### 3.2. DBM excited state dynamics of relaxation and TA target analysis.

Direct designation of the spectral features in the TA spectra cannot usually be done based on a single experiment. The investigation of the excited state dynamics of the molecules is based on a comparison of the TA data obtained from different samples under different conditions such as environment, temperature etc. Moreover, composite structures such as the compounds under studymay inherit some properties from their component elements, which means knowledge about those properties is useful for the characterization of the system itself. Following this logic, prior to the discussion of the excited state dynamics of the europium complexes, we present the data of TA analysis of the DBM ligand.

Transient absorption spectra of DBM and representative spectra as well as fitting of the kinetic traces comes after the target analysis of the transient data thatuse the scheme presented on CHART 4, are plotted on Figure 2. The negative band at 343 nm on the transient map is the ground state bleach (GSB), which usually has the shape of the ground state absorption band, but here it is strongly affected by the pump pulse artefacts at 350 nm. The other negative band at 450 nm in the TA spectrum is the stimulated emission (SE), which doesn't match completely steady state spontaneous emission because ofthe superposition of the excited state absorption (ESA) bands at 400, 600 and 720 nm in this differential spectrum (Figure 2c).

Time evolutions of these bands,positive and negative, present the dynamics of the relaxation of the excited state of DBM. In this case the decay of the bands is four-exponential. We obtained the best fit of the experimental data using the branched scheme with five compartments as target model (CHART 4).

In this paper we use the term 'compartment' to denote particular species from the kinetic scheme used for the target analysis. These species or compartments may represent a real electronic state such as $S_1$ or $T_1$ states or a distinct transient species from a given electronic state such as a molecule in an excited singlet state right after excitation or in the relaxed solvation state, both of the same $S_1$ electronic state, for example. This terminology is already accepted in literature[29].

The first three lifetime constants in the model present the evolution of the first excited singlet state. The fourth lifetime constant presents relaxation of the triplet state. The fifth decay constant was kept fixed to 20 ns and was used as an infinite time component accounting for the spectral feature at the GSB area. Looking at the TA spectra one can find that there is a significant negative signal long after the excited state relaxation. This is due to a photochemical reaction that transforms the excited molecules to some other species in a different ground state conformation than the initially excited DBM molecules. As has been discussed before[39, 40], this is the mostlya non-chelated enol (NCE) form of DBM given that the efficiency for photoketonization of DBM in protic solvents is negligible. These species have different ground state absorption compared to the chelated enol and this causes a remaining negative signal as an infinite component on the TA experiment time scale. The remaining signal at GSB area is due to non-chelated enol formation at ca. 30% from initial excitations, which is in agreement with previous studies[40]. This feature is characteristic of TA spectra of the free DBM ligand and is missing in TA spectra of the DBM complexes where coordinated chelate saves configuration during the excited state relaxation. The photoproduct is unstable and it returns to the ground state on the ms time scale[40], which keeps its concentration low at the photo-stationary state at 1 kHz repetition rare of the Ti-sapphire laser and continuous shifting of the sample.

**CHART 4. Excited state dynamics scheme of DBM used for target analysis. The lifetime constants are associated to the compartments in target analysis model used to describe the excited states and dynamics of relaxation of the excited molecules.**

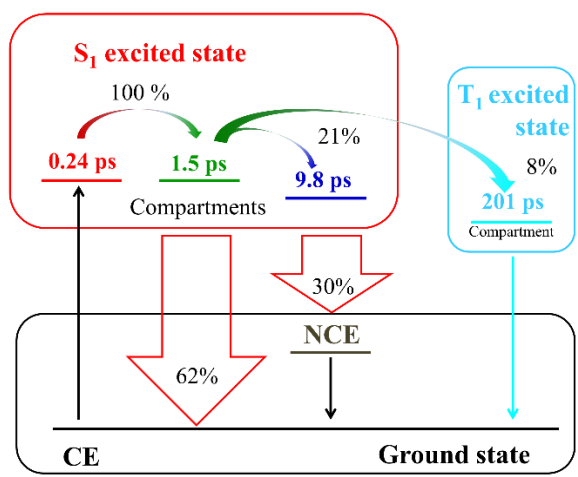

The time evolution of the transient spectra of DBM presents dynamics of relaxation of its singlet and triplet excited states. The decay is fitted with four exponents and regression analysis gives the species associated spectra (SAS) (Figure 2a) thatcan belinked withthe corresponding time constants (Table 2). The first sub ps state is attributed to the Franck-Condon (FC) state. In this state an excited molecule posse's geometry of the ground state and the excitation is achieved only by active vibronic transitions.

One characteristic feature for this state is SE where the coupling with the ground state through active vibronic modes is still strong. Moreover, it possesses two ESA bands. The long wavelength ESA band is centred at 720 nm and the other band is partly overlapped by GSB causing a maximum in differential absorption spectra around 400 nm. The lifetime of this state is 0.23 ps. In the next state the molecule entirely loses the SE spectral feature and the EAS band centred at 720 nm is significantly less intense. This excited state species lives <2 ps and it is coupled with two other states the last singlet compartment and triplet state, respectively. The ESA band which is centred at 600 nm and at maximum intensity after 4 ps we treat as absorption band of the first excited triplet state. The transfer of excitations amounts to 8% of all initially populated singlet states. Although this intersystem crossing (ISC) is not very efficient, it is an example of a very fast one with rates as fast as $1.3\times10^{11}$ s$^{-1}$. The lifetime of the triplet state is around 200 ps.

The target analysis shows that ISC along with internal conversion competes with other processes where the molecule suffers significant conformational changes extending to the last singlet compartment. We assume these are probably out-of-plane twistings of the chelate ring around the double bond. The spectral feature of this state is one ESA band with a maximum below 400 nm and a lifetime below 10 ps. The reason for this fast decay may be conical intersection with the ground state hyper-surface. After analysis of the TA data we did not find evidence for triplet state population from this state.

For the calculation of the branching ratios between compartments in a model for target analysis such as the one in CHART 4 the kinetic traces from the GSB band should be used. To calculate these ratios in our experiments we used only kinetic traces at the shoulders of the GSB band because the GSB minimum matches the excitation wavelength and this area contains a lot of optical artefacts cased by scattering of the pump pulse, which makes this data unusable. For clarity in the most of the figures in this paper presenting TA data, the data points around the excitation wavelength are deleted from the transient map. We assume that the signal below 335 nm is only a result of the bleaching of the ground state absorption of DBM molecules and reflects the dynamics of recovery of the molecules in the ground state that come from any excited state compartment. We use this area of the spectrum to calculate the branching ratios. For this purpose we keep fixed the decay constants of the transitions between compartments, which are already determined from the fits of the kinetic traces of the all other bands in the TA spectrum and only the pre-exponential factors in the kinetic equations are optimized during the fit iteration. These pre-exponential factors contain the branching ratio constants and the constants which are a product of the molar absorptivity. The latter are kept as a shared global fit parameter during iteration because they account for the molar absorptivity of the molecules in ground state. In this way we overcome the dependence between the fit parameters and the fit can converge to a minimum. We use the same approach to find the branching ratios for the rest of the samples studied here.

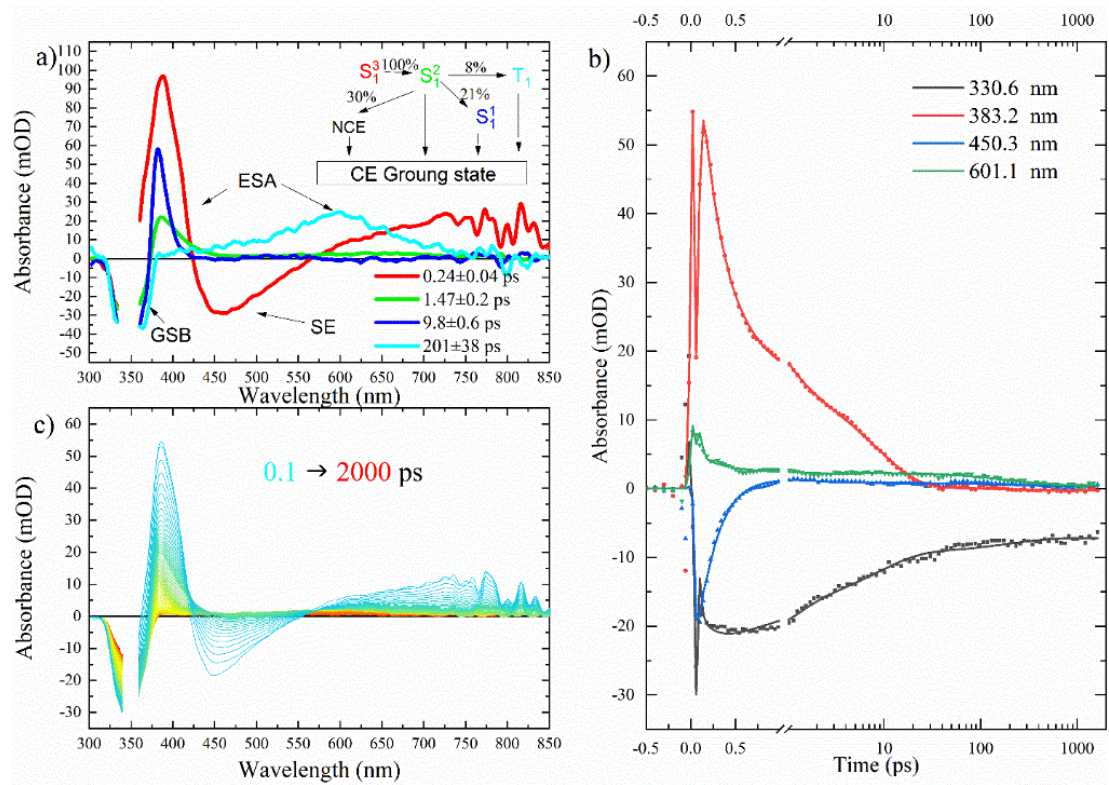

Figure 2. Transient absorption of DBM in ethanol at 23°C pumped at 350 nm. a) is species associated spectra (SAS) and b) – fits of kinetic traces at specific wavelengths comes after target analysis of transient matrix –c). The deleted area around 350 nm contains pump pulse artifacts.

**Table 2** Lifetimes and transition probabilities of the excited states come after target analysis of the TA spectra of the compounds in different solutions.

| | $S_1^3$ (ps) | | $S_1^2$ (ps) | | $S_1^1$ (ps) | | $T_1$ (ps) | $\eta_{ISC}$ |
|---|---|---|---|---|---|---|---|---|
| DBM Ethanol | 0.24±0.04 ps | 100% → | 1.5±0.2 ps | 8% 21% | 9.8±0.6 ps | → | 201±38 ps 30% | 8% |
| DBM pH11 Ethanol | 0.15±0.04 ps | 100% → | 1.2±0.1 ps | 7% 29% | 9.3±0.6 ps | → | 189±25 ps 30% | 7% |
| Eu(DBM)$_3$.H$_2$O Ethanol | 0.43±0.01 ps | 72% → | 18.9±0.3 ps | 20% → | 156±24 ps | 0% → | | 0% |
| Eu(DBM)$_3$.EDA Acetonitrile | 0.34±0.03 ps | 77% → | 16.5±0.3 ps | 41% → | 151±10.8 ps | 100% → | 471±57 ps | 32% |
| Eu(DBM)$_3$.Phen Acetonitrile | 0.23±0.01 ps | 81% → | 11±0.9 ps | 50% → | 64±9.2 ps | 59% → | 218±16 ps | 20% |
| Eu(DBM)$_3$.Phen Ethanol | 0.58±0.02 ps | 90% → | 4.9±0.2 ps | 43% → | 36±1.3 ps | 38% → | 227±7 ps | 9% |
| Al(DBM)$_3$ Acetonitrile | 1.79±0.04 ps | 100% → | 8.5±0.7 ps | 43% → | 75±2.3 ps | 0% → | | 0% |
| Al(DBM)$_3$ PMMA | 8.0±0.56 ps | 46% → | 45±5 ps | 59% → | 412±117 ps | 25% → | Inf | 7% |

$\eta_{ISC}$ is the efficiency for ISC to triplet state. The standard error of the branching ratios is less than 15% of the value.

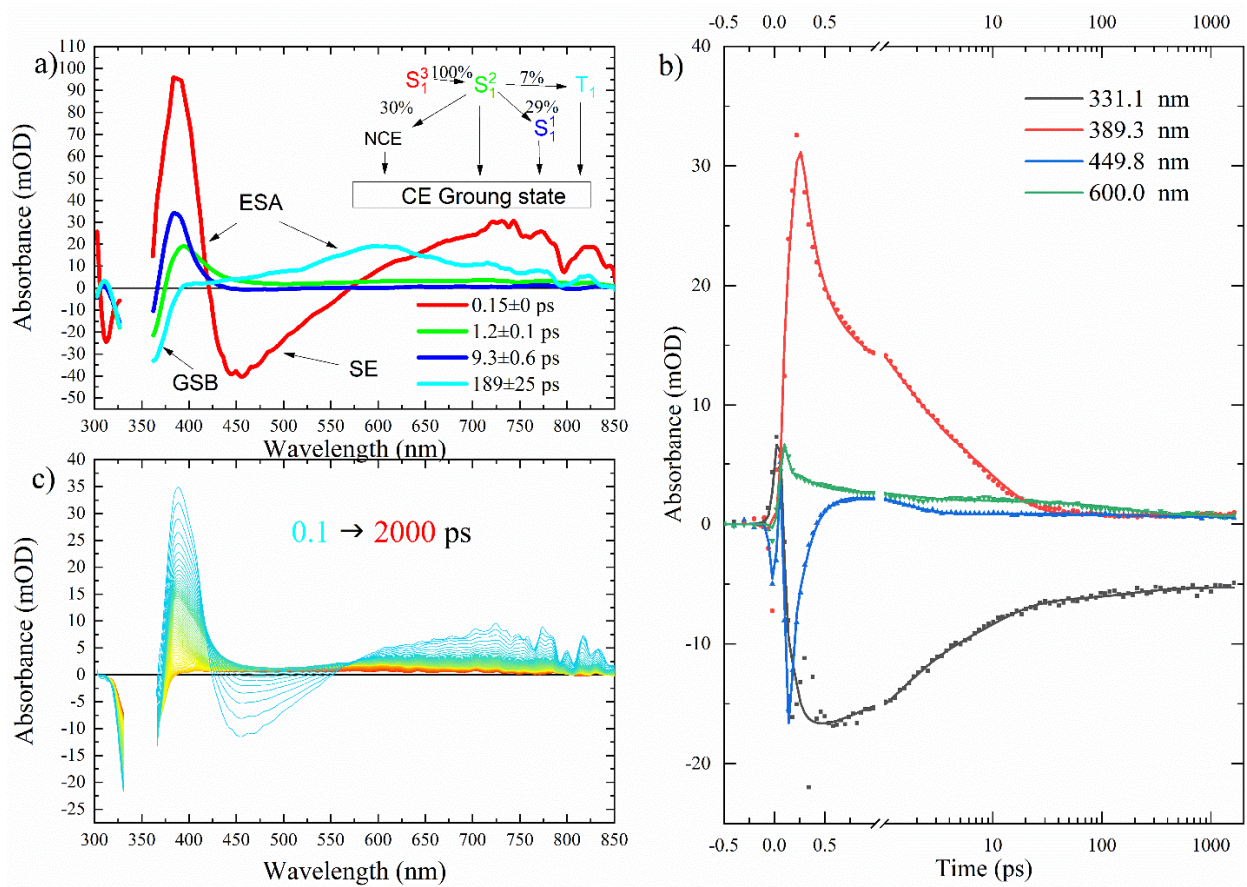

Figure 3. Transient absorption of (DBM)⁻ in ethanol pH 11 at 23 °C pumped at 350 nm. a) is species associated spectra (SAS) and b) – fits of kinetic traces at specific wavelengths comes after target analysis of transient matrix –c). The data around 350 nm is deleted because contains pump pulse artifacts.

As can be seen from Figure 2, the DBM has extremely fast relaxation dynamics and a short lifetime of the excited state. Such dynamics are usually characteristic of molecules that undergo significant structural changes on the potential surface of the excited state. As a result, the minimum energy of the excited state relative to some of the vibrational coordinates is significantly shifted from that of the ground state.This is probably the reason for the rapid relaxation of the excited state.It might due to the effective coupling with the higher vibration levels of the ground state which takes place and which accelerates the internal conversion or causes conical intersections between the surfaces.Such a structural change may be proton transfer, as previously suggested[19, 20].Due to the relative displacement of the two potential surfaces, the vertical absorption transition results in a Franck–Condon state located at higher vibrational levels. The vibrational relaxation of this state usually takes hundreds of fs. This process is followed or accompanied by slower intramolecular vibration redistribution (IVR).

It was interesting for us to check excited state dynamics of deprotonated DBM. We record the TA spectra of DBM in alcoholic solution at pH 11 (Figure 3) where there isno hydroxyl hydrogen bonded to the DBM molecule. Surprisingly, this data doesn't show significant differences in comparison with the TA data for DBM in a neutral solution.The decay pattern and the spectra are virtually the same. However, there is adifference for the rate constant values measured for these two samples. In the case of deprotonated DBM the values of decay time constants other than the one for the first compartment appear to be a little bit longer. It might be due to the absence of H—O stretching high energy vibrational mode, which facilitates internal conversion. At the same time, there is a reverse tendency for the shortest decay constant of the first compartment. We measure 240 fs and 150 fs decay constants for DBM and (DBM)$^-$, respectively. Given that the response function of our instrument is around 130 fs and the signal is convoluted with the pump pulse, interpretation of these values must be done carefully.

According to these observations and preserving the decay pattern going from DBM to (DBM)$^-$ we cannot accept the idea of excited state proton transfer in its classical meaning. Rather we assume that this fast dynamics is due tothe redistribution of the uniform π-electron density of the chelate in ground state to enol or diketo-like geometry in the excited state where rotation around apha bond is easier. Diketo-like geometry increase (n,π) character of the excited state where ISC to triplet state is more probable and coupling with the ground state, respectively the emission are restricted. Such a time-dependent conformational transformation can block some relaxation pathways putting the system in new compartment with a loner decay time.

## Excited state dynamics of the complexes.

The TA spectrum of of Eu(DBM)$_3$.H$_2$O complex in ethanol is presented inFigure 4. As the figure shows it contains the same basic spectral features as those of the DBM ligand, but nothing except the positions of the bands is the same. Both excited state lifetime constants and the spectral features in the TA mapsuch as the number, shape and relative intensity of the ESA bands of the complex are different.By comparingFigure 4 and Figure 2 one can see thatthe band at 720 nm is much more intense in the TA spectra of the europium complexes than in spectra of the DBM ligand. This means that in this case coupling between first and the second singlet excited state,which is accessible from the ground state through a 260 nm absorption band excitation, is stronger. Moreover, there isfull recovery of the GSB area after the antenna's excited state relaxation, whichis evident for all of the DBM complexes studied here.The most likely reason for this is coordination with ametal ion that saves the chelate structure upon excitation. The time evolution of the TA spectrum oftheEu(DBM)$_3$.H$_2$O complex (Figure 4) shows that the ESA band at 720 nm remains until the end of excited state relaxation.Immediately after excitation the intensity of the ESA bands centred at 720 and 400 nm is almost equal. During excited state evolution there is a change in the relative intensities of these bands and as a result the red ESA band becomes more intense. It is noteworthy that the position of these bands during relaxation remains constant. This means that the energy of the excited state remains constant or at least the energy difference between first excited and upper excited states stays unchanged. Most probably this observation is the case where is monitored evolution of single electronic excited state.

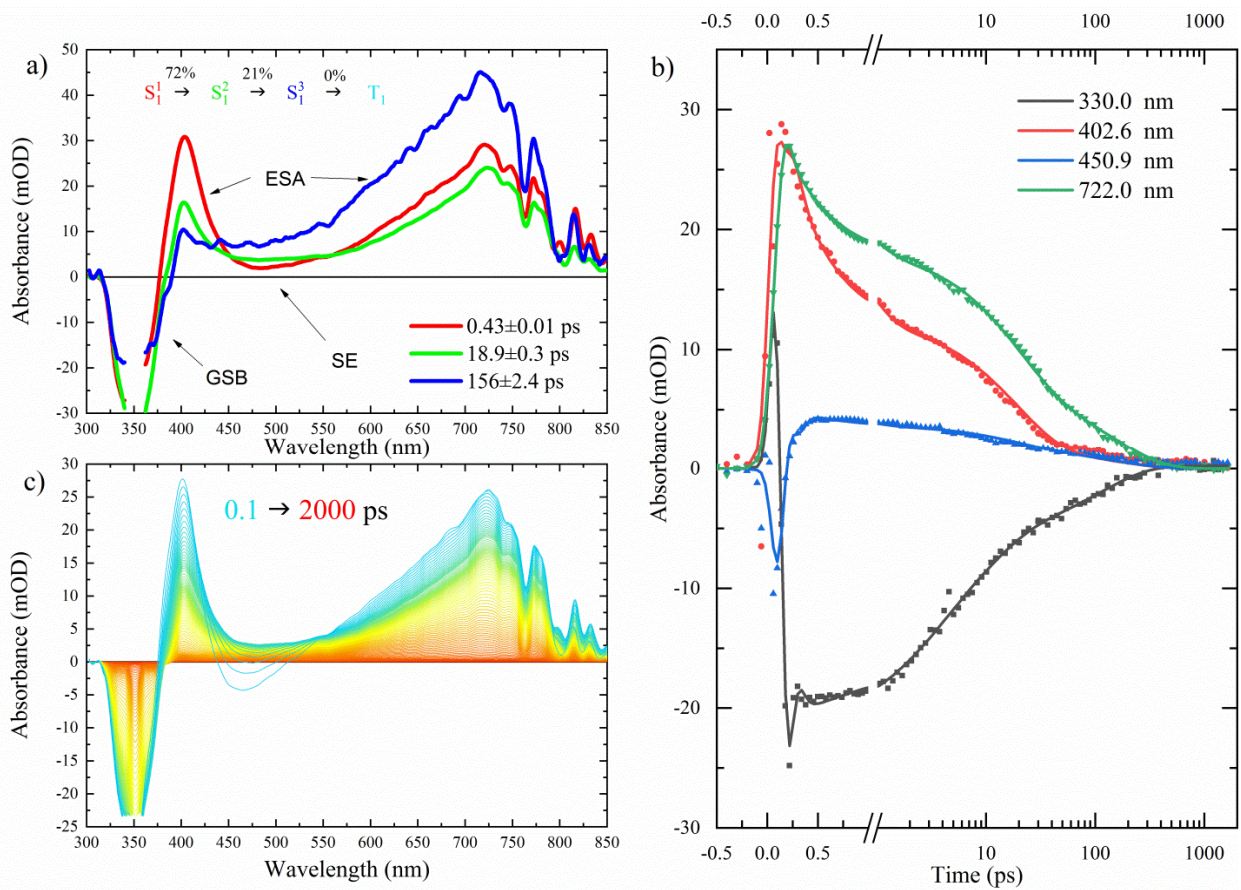

Figure 4. Transient absorption of Eu(DBM)$_3$.H$_2$O in ethanol at 23 °C pumped at 350 nm. a) is evolution associated spectra (EAS) and b) – representative fits of kinetic traces come after target analysis of transient matrix –c). The area around 350 nm contains pump pulse artifacts.

The best fit of the TA data of Eu(DBM)$_3$.H$_2$O complex we obtain using the model for target analysis present as inset in Figure 4a. This scheme presents a sequential transfer of excitations through compartments and also accounts for the direct relaxation to the ground state that comes from each compartment. For simplicity, we have omittedthe arrows for the pathways of direct relaxation in the chart buthave included the transition probabilities between compartments as branching ratios.

The evolution associated spectra (EAS)we obtained are plotted on (Figure 4a). One can see that these spectral features are almost identical and they can hardlybe referred to different electronic excited states. The difference in EAS going through compartments is in the relative intensities of both ESA bands, suggesting that there is a change in transition probabilities rather than their energies. We assume that these ESA bands represent the first excited singlet state. The relaxation of this excited state manifest three-exponential decay. We believe that this three-exponential behaviour of the excited state decay does not represent actual physical states of the molecule. Rather it isan effect of intramolecular vibrational redistribution (IVR). Due to femtosecond time resolution of the excitation pulse its spectral band range is wide enough to produce a coherent superposition of various excited states. Assuming the sequential coupling model[41-44]each one of these states is coupled with another, so-called dark state and finally with a dense set of states resulting in a decay curve which can best be approximated to a sum of exponents. This behaviourin energy dissipation is characteristic of systems that have excited state relaxation time similar tothe time for IVR between states differently coupled with the ground electronic state. As it will become clear later (see below) this behaviourin energy dissipation is the same for all complexes of DBM.

However, there is a question about the triplet state and the mechanism for energy transfer. The fact that Eu(DBM)$_3$.H$_2$O complex shows some luminescence from the $^5D_0$ excited state even though the quantum yield is less than 1% (Table 1) and there is absence of the triplet ESA band in the TA spectrum can be explained as follows: either the population of that state is too low and below the dynamic range of our transient spectrophotometer, or there is no triplet state population at all, but some energy transfer to the excited states of the Eu(III) ion goes through the antenna's excited singlet. These statements are both possible simultaneously.

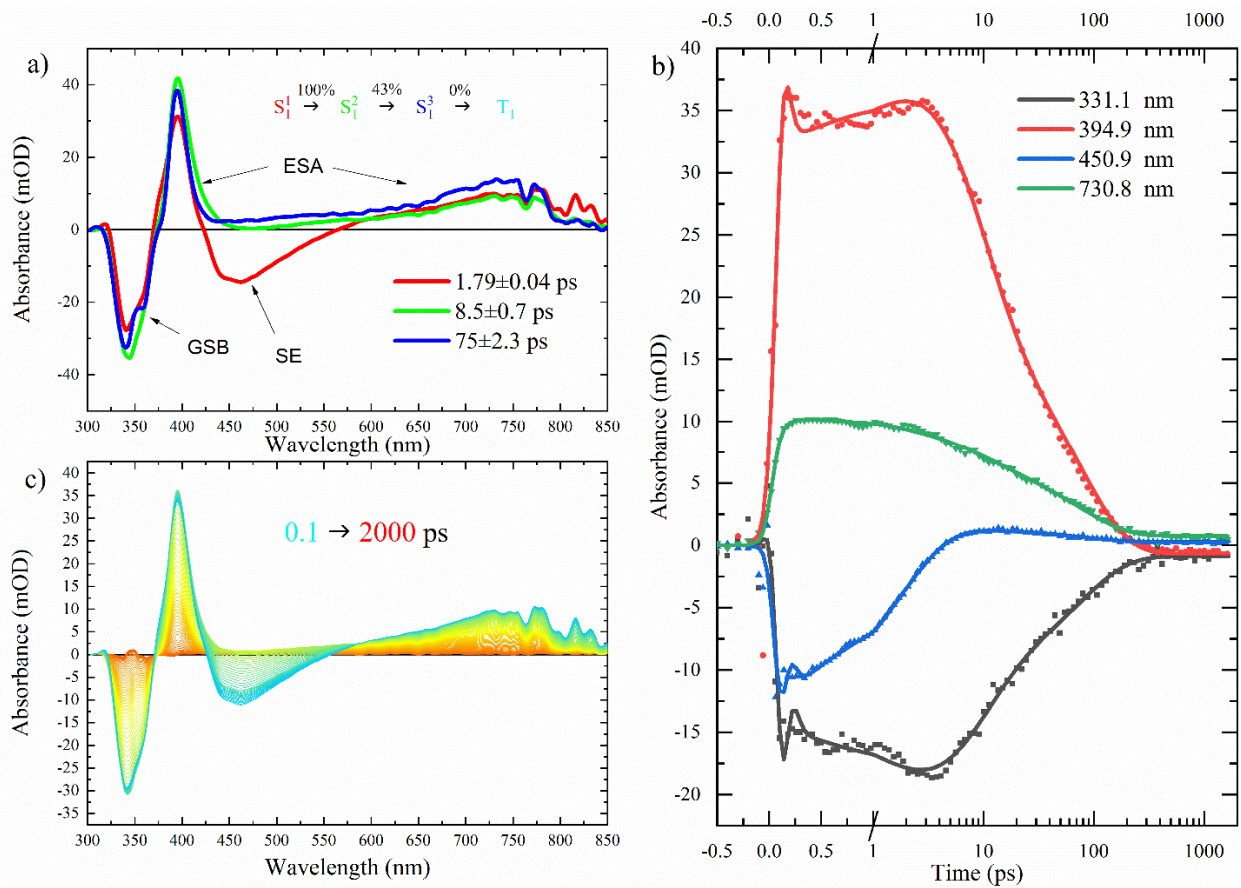

Figure 5. Transient absorption of Al(DBM)$_3$ in acetonitrile at 23 °C pumped at 350 nm. a) is evolution associated spectra (EAS) and b) – representative fits of kinetic traces come after target analysis of transient matrix –c). The area around 350 nm contains pump pulse artifacts.

We have measured the TA of the aluminium complex of DBM to check if this three-exponential decay pattern of the singlet state in Eu(III) complex is an intrinsic property inherited by the DBM ligandor the dynamics depends on the energy transfer. Given that Al(III) has different energy levels compared to Eu(III) a different relaxation mechanism can be expected.

The results are plotted in Figure 5. As shown, the dynamics of relaxation of Al(DBM)$_3$ complex is also demonstrated by three-exponential decay. In this case there are even more similarities between representative spectra of the compartments. Here the SE manifests not only in the first compartment but also in the second. This supports the hypothesis that at least two compartments havea singlet nature. The TA data of the Al(DBM)$_3$ complex suggest a prolonged lifetime of its first compartment. Moreover, TA spectra of the complex look like spectra of the free DBM more than those of the Eu(DBM)$_3$.H$_2$O complex. Indeed the long wavelength band centred at 750 nm does not disappear until the end of the relaxation but it is less intense compared to the band at 395 nm, which is characteristic ofthe spectra of DBM. Apart from the visual similarity of the TA spectra, the Al(DBM)$_3$ complex has nothing in common with the dynamics of DBM. The ESA band for the triplet state is missing and there is no evidence of significant structural changes duringthe excited state relaxation of the aluminium complex. This compound is photostable upon photoexcitation and the GSB recovery is full after relaxation. All these features are the same as those oftheEu(DBM)$_3$.H$_2$O complex. The results show thatthe dynamics of relaxation and particularly the lifetime of the third compartment of the Al(DBM)$_3$ complex is less than that of the Eu(DBM)$_3$.H$_2$O complex. This is unexpected given that this is a complex where energy transfer from ligand to metal ion has notbeen reported to the best of our knowledge. We may look for the answer in the assumption that the aluminium complex shares some relaxation pathways typical of the DBM ligand, which may be supported by the visual similarity of the TA spectra of both compounds.

There is still a question about the nature of the third compartment, but the visual similarity of representative spectra (EAS) for the compartments and the shorter lifetime of the third compartment measured at the Al(DBM)$_3$ complex compared to the Eu(DBM)$_3$.H$_2$O complex convinces us that the third EAS is not the spectrum of the triplet state.

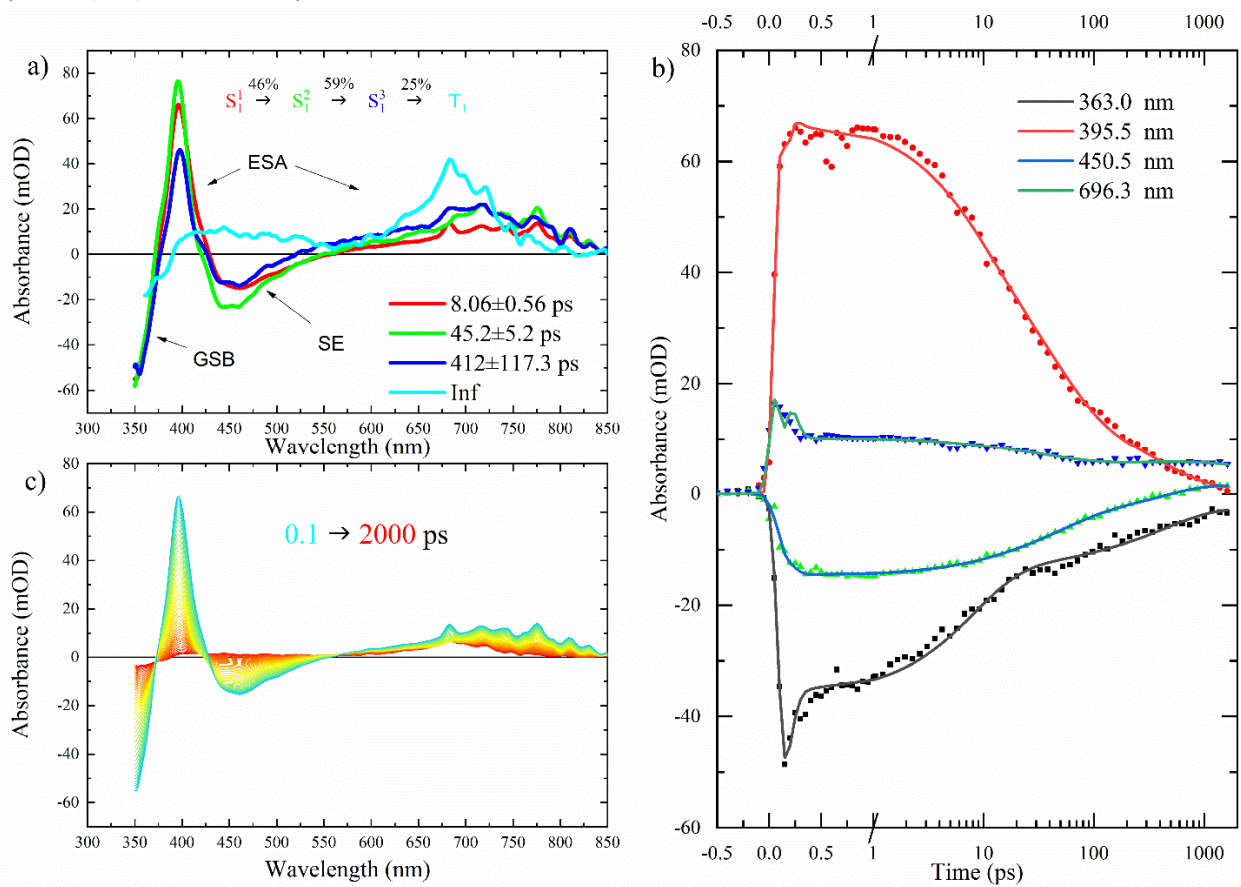

Figure 6. Transient absorption of Al(DBM)$_3$ in PMMA matrix at 23°C pumped at 340 nm. a) is evolution associated spectra (EAS) and b) – representative fits of kinetic traces come after target analysis of transient matrix –c). The area below 350 nm contains pump pulse artifacts.

Nevertheless, the question about the triplet state remains. To find the answer, we needed a sample to check where we can detect the existence of the triplet state. To do this, we prepared polymer films doped with the compounds under study and coated over glass plate. Fortunately, the samples containing DBM ligand and Al(DBM)$_3$ complex showed phosphorescence at room temperature. The fluorescence and phosphorescence excitation and emission spectra of Al(DBM)$_3$ complex in the polymethylmethacrylate (PMMA) matrix are presented on Fig. S11 and Fig. S12, respectively. As shown, the quantum yield of the phosphorescence is higher under 260 nm excitation and even in this hard sample coupling of the first singlet state (accessible through 350 nm transition) with the triplet state is weak. However, the phosphorescence excitation spectrum clearly shows a visible phosphorescent signal under excitation at 340 nm, which proves the existence of the triplet state.

The results from the analysis of the TA data for the Al(DBM)$_3$ complex in PMMA were more than surprising and are shown on Figure 6. First, the existence of the triplet state is confirmed by the TA where it appears as an infinite spectral feature in the TA map at 680 nm (Figure 6c). Second, the dynamics of relaxation of the complex in the polymer matrix also shows three-exponential decay. The first three compartments certainly have a singlet nature. Their EAS are almost identical and all three show SE. Moreover, the lifetime constants of all three singlet compartments of the Al(DBM)$_3$ complex in the polymer matrix are increased proportionally ca. 5-6 times compared to those in solution. This means that the energy dissipation mechanism is an intrinsic property of these molecules and the nature of the excited state and number of the compartments do not depend on the environment. The proportional increase in the lifetimes of all compartments is possible if they belong to a single electronic excited state whose vibrational relaxation is blocked in the polymer cage. The lack of vibrations also reduces the efficiency of the transition between compartments (Figure 6a). Usually the transition efficiency between the first and second compartments is nearly 100% (see Table 2). In this case it is below 50%. This gives less than 7% efficiency for ISC. However, this is enough to record and analyse the ESA for the triplet state. Two spectral features distinguish the EAS of the triplet state from these of compartment of the singlet state. First, there is a 40 nm hypsochromic shift of the absorption maximum of the red band and second, the blue ESA band from singlet state spectra is missing. Recording the ESA for the triplet state and assigning the first three compartments to singlet state in this sample helped us to analyse the excited state dynamics of the Eu(III) complexes.

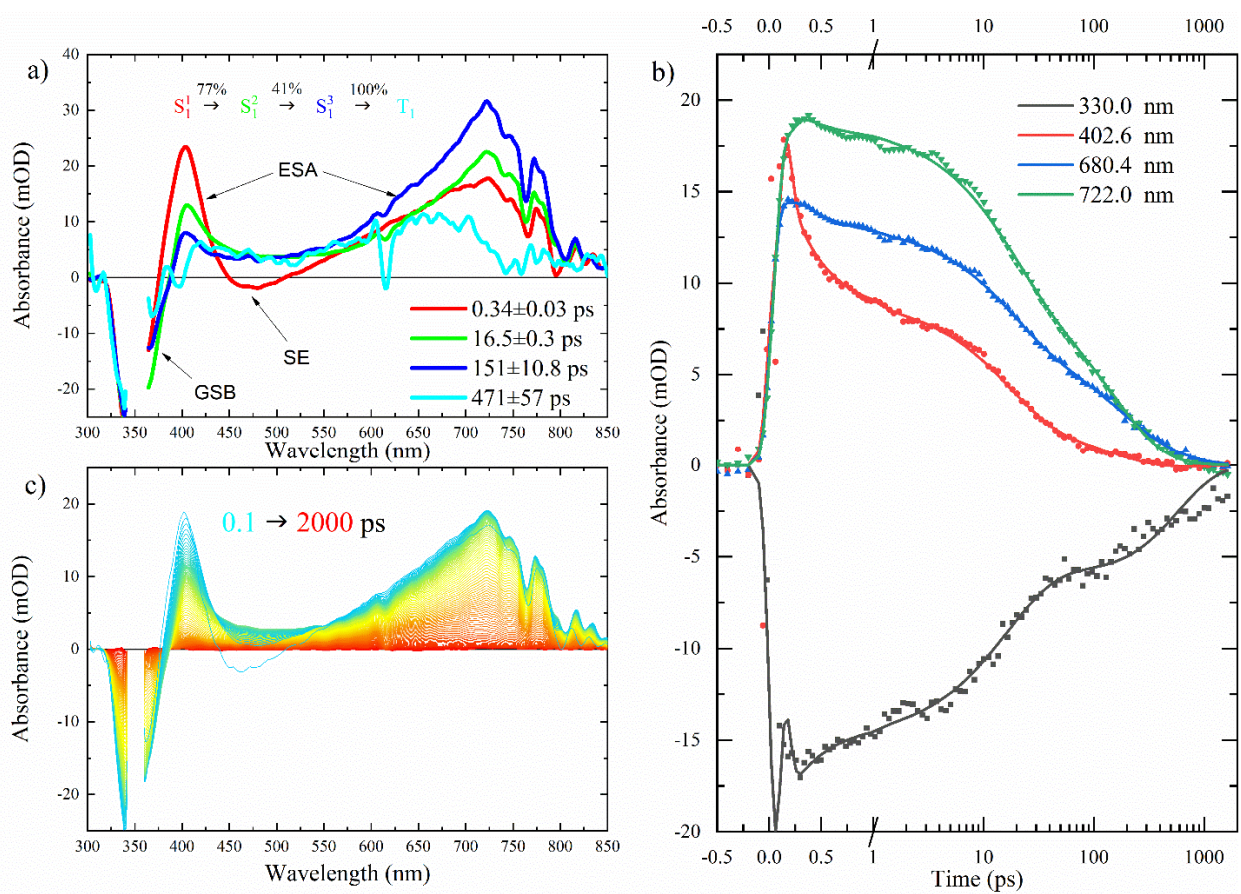

Figure 7. Transient absorption of Eu(DBM)$_3$.EDA in acetonitrile at 23°C pumped at 350 nm. a) is evolution associated spectra (EAS) and b) – representative fits of kinetic traces come after target analysis of transient matrix –c). The area around 350 nm contains pump pulse artifacts and the spectral artifacts above 750 nm are due to pure white light quality around laser wavelength at 795 nm.

Analysing the TA data for the Eu(DBM)$_3$.EDA and Eu(DBM)$_3$.Phen complexes in acetonitrile solutions (Figure 7, Figure 8) reveals the reason for luminescent enhancement in the Eu(III) DBM complexes caused by the replacement of the water molecule from the inner coordination sphere of the complex with a nitrogen-containing ligand. As shown on Figure 7 and Figure 8 the existence of a fourth compartment that bears the spectral features of the triplet state is evident in the TA spectra of these two complexes. The EAS spectrum of the triplet state has one TA band with a maximum at 680 nm. The lifetime of the triplet state is around 470 ps and 220 ps for Eu(DBM)$_3$.EDA and Eu(DBM)$_3$.Phen complexes, respectively. As can be seen Table 2 in the same tendency is valid for the decay constants of the other compartments and the Eu(DBM)$_3$.Phen complex shows slightly faster dynamics compared to the Eu(DBM)$_3$.EDA complex.

We have used the kinetic traces from the GSB area of the TA spectrum to calculate the branching ratios and the efficiency for ISC ($\eta_{ISC}$, Table 2) to triplet state, respectively. The results are in good agreement with calculated values for energy transfer efficiency to the Eu(III) ion excited states ($Q_{Sens}$, Table 1). These values are almost identical, which means the energy transfer is nearly 100% efficient. This is not surprising because it is evident for other europium 1,3-diketonates.

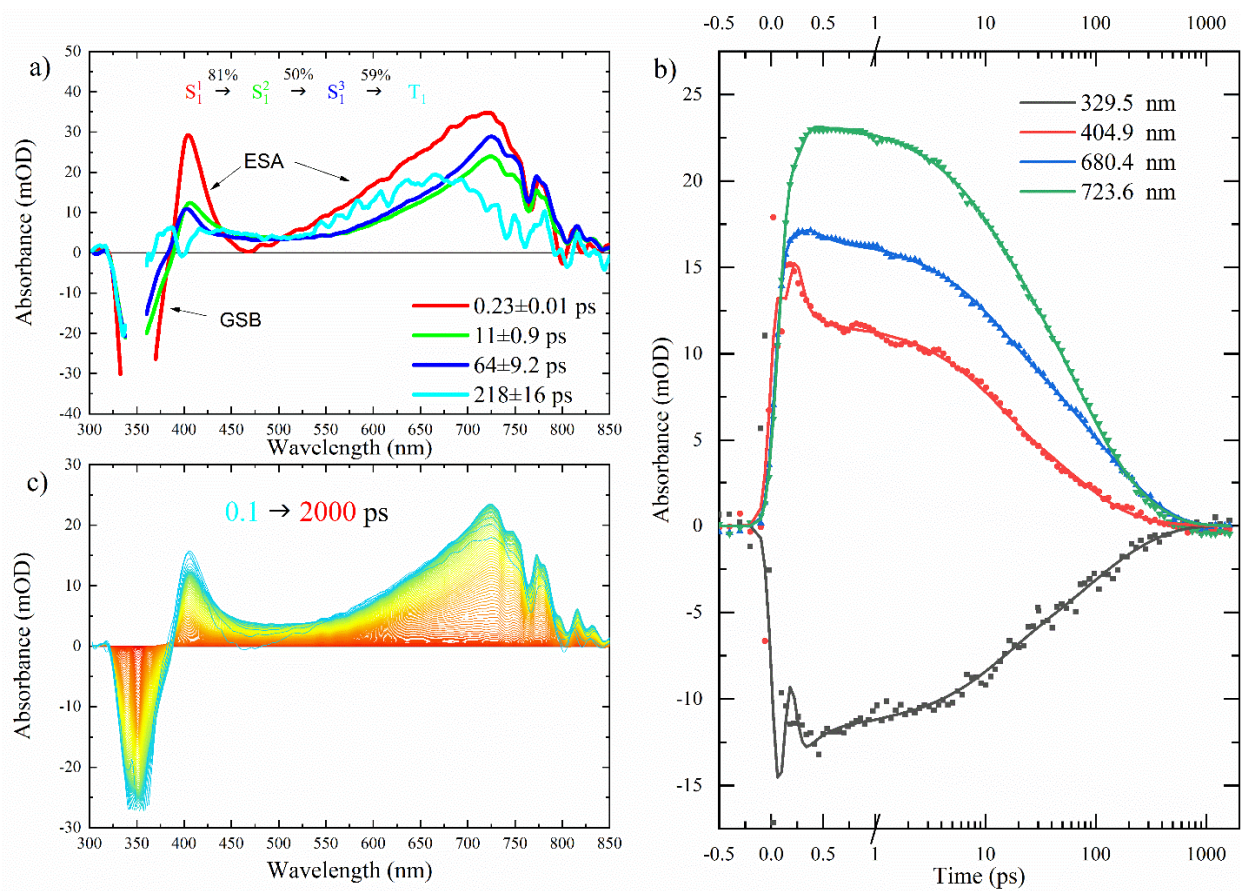

Figure 8. Transient absorption of Eu(DBM)$_3$.Phen in acetonitrile at 23 °C pumped at 350 nm. a) is evolution associated spectra (EAS) and b) – representative fits of kinetic traces come after target analysis of transient matrix –c). The area around 350 nm contains pump pulse artifacts and the spectral artifacts above 750 nm are due to pure white light quality around laser wavelength at 795 nm.

In case of the Eu(DBM)$_3$.EDA complex ISC goes most efficiently from the third singlet compartment. As Figure 7shown on all excitations from this compartment go to triplet state. As a result the recovery of the GSB with a 151 ps time constant does notmanifest in kinetic traces at this area. If we accept that the main donor for triplet state excitations is the third compartment of the singlet state,it can explain the lack of luminescent efficiency in the Eu(DBM)$_3$.H$_2$O complex. Comparing the transient map of the all Eu(DBM)$_3$.L complexes beyond spectral and dynamic similarities one can see that in the case of Eu(DBM)$_3$.H$_2$O complex more than 85% of all excitations are lost during the lifetime of the first two compartments of the singlet state. The lack of population of the third compartment reduces the chance for triplet state population. The results show that replacing the auxiliary ligand does not change the number of the time constantswhich presents the mechanism of relaxation in the complex, but rather it influences the relative distribution of the excitations over the compartments and hence the efficiency for ISC. If we assume that this distribution depends onthe coupling strength between these quantum states then the nature of the compartments is crucial. Introducing the Lewis bases in the inner coordination sphere of the Eu(DBM)$_3$.EDA and Eu(DBM)$_3$.Phen complexes increases electron density into the chelate, which may induce (n, π*) character of the singlet excited state where ISC is more efficient in general. On the other hand, if these compartments correspond to some conformational states of the molecule then changing the ligand may displace the relative position of the potential minimum of these states, which may affect the energy barrier (or coupling) between them and hence the relative distribution of the excitations. Unfortunately, we don't know the nature of this phenomenon. Clarifying this certainly demands theoretical investigation. What becomes clear from this study is the correlation between the luminescent quantum yield andthe transition efficiency between singlet compartments and the triplet state, respectively.

## 4. Conclusions

The aim of this study was to expose the reason for the luminescent enhancement caused by replacement of the water molecule fromthe inner coordination sphere of the Eu(DBM)$_3$.H$_2$O complex with ammonia, amines or heterocyclic nitrogen containing compounds, using transient absorption spectroscopy.

We compare TA spectra of several compounds starting witha DBM free ligand and proceed with its aluminium and europium complexes in different solvents, pH and polymer matrices. The results for all samples show constant three-exponential decay of the singlet excited state presented by three compartments.We conclude that this feature is inherited from DBM to its complexes and suggests that this decay pattern is anintrinsic property of the ligand and does not depend on the surrounding.

Investigation of the TA dynamics of the complexes with nitrogen containing ligands shows a clear correlation between energy transfer efficiency and the population of the triplet state. Replacing the water molecule from the Eu(DBM)$_3$.H$_2$O complex with a nitrogen-containing ligand acts as a "switch" turningonan ISC as effective channel for energy transfer toward Eu(III) ion.

Moreover, after analysingthe TA data we conclude that ligand exchange not only enables ISC, but also changes relative transition efficiency and branching ratios between the compartments of the singlet excited state. We found that the high population of the third singlet compartmentis essential for the triplet state occupation. In the case ofthe Eu(DBM)$_3$.H$_2$O complex 85% of all excitations are lost in the first two compartments.Although there is some population of the third singlet compartmentof this compound, by analyzing its TA spectrum we did not find evidence for the triplet state existence.

## Conflicts of interest

There are no conflicts to declare.

## Acknowledgements

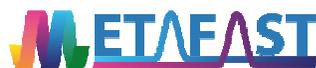

This project has received funding from the European Union's Horizon 2020 research and innovation programme under grant agreement No.899673.

This work reflects only author view and the Commission is not responsible for any use that may be made of the information it contains. Art.29.5 GA

## Associated content

### Supporting information

There is additional supporting material containing NMR, IR, absorption and luminescent excitation and emission spectra of the compounds under study. This material is available free of charge online.

## Author information

### Corresponding Author

sstanimirov@chem.uni-sofia.bg

### Present Address

Sofia University, Faculty of Chemistry and Pharmacy 1 James Bourchier blvd., Sofia 1164, Bulgaria.

## Refferences:

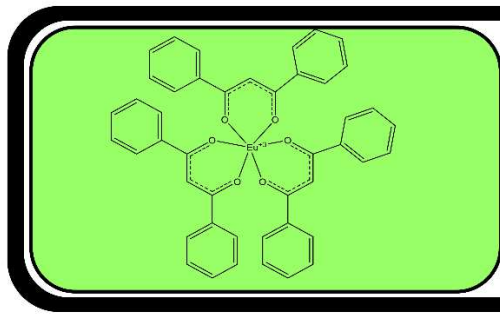

Graphical abstract